# Single-Crystal Organic Field Effect Transistors with the Hole Mobility ~ 8 cm$^2$/Vs

V. Podzorov [a)], S. E. Sysoev, E. Loginova, V. M. Pudalov [b)], and M. E. Gershenson
*Department of Physics and Astronomy, Rutgers University, Piscataway, New Jersey 08854*

We report on the fabrication and characterization of *single-crystal* organic p-type field-effect transistors (OFETs) with the field-effect hole mobility $\mu$ ~ 8 cm$^2$/V·s, substantially higher than that observed in thin-film OFETs. The single-crystal devices compare favorably with thin-film OFETs not only in this respect: the mobility for the single-crystal devices is nearly independent of the gate voltage and the field effect onset is very sharp. Subthreshold slope as small as $S$ = 0.85 V/decade has been observed for a gate insulator capacitance $C_i$ = 2 ± 0.2 nF/cm$^2$. This corresponds to the *intrinsic* subthreshold slope $S_i \equiv SC_i$ at least one order of magnitude smaller than that for the best thin-film OFETs and amorphous hydrogenated silicon ($\alpha$-Si:H) devices.

The quest for high-performance organic field-effect transistors (OFETs) has resulted in a significant increase of the charge carrier mobility $\mu$.[1] In the best devices based on thin organic films, values of $\mu$ up to ~ 1.5 cm$^2$/V·s have been reported.[2] This performance is already comparable with that of amorphous-silicon FETs.[3] However, there are still several important issues to be resolved, most of them being associated with grain boundaries and interfacial disorder in organic thin films. Indeed, currently these structural defects are the major factor which limits the mobility,[2,4] causes the dependence of the mobility on the gate voltage,[5,6] and results in the broadening of the on/off transition.[7] Grain boundaries can be eliminated in devices fabricated on single crystals of organic semiconductors, which enables to explore the role of other factors.

Recently, we developed a technique for the fabrication of single-crystal OFETs,[8] which allowed us to completely eliminate the inter-crystalline boundaries. In this Letter, we report on the optimization of this technique, which has resulted in a dramatic increase of the field effect hole mobility up to $\mu$ ~ 8 cm$^2$/V·s. The large magnitude of $\mu$ is not the only advantage of the single-crystal devices: their mobility is nearly gate-voltage independent, and the onset of conductivity is very sharp. Comparison between single-crystal and thin-film OFETs helps to identify the characteristics of the latter devices, which are associated with structural defects.

High-quality rubrene crystals have been grown from the vapor phase in a stream of ultra-high-purity hydrogen in a horizontal reactor.[9] Several key factors affect the crystal quality. One of the important parameters is the difference in temperature between the sublimation zone and the growth zone, $\Delta T = T_{sblm} - T_{growth}$, which is an analog of the *supersaturation* at thermal equilibrium. The regime of small supersaturation, when $T_{sblm}$ is set close to the sublimation threshold of rubrene, is crucial for the mobility improvement. The crystal growth in this regime proceeds by the flow of steps at a very low rate ($\leq 5\times10^{-7}$ cm/s in a direction perpendicular to the facet), and results in a flat surface with a low density of growth steps.[10] In our experiment, it took up to 50 hours at $T_{sblm}$ = 300 $^{\circ}$C and a H$_2$ flow rate 100 cc/min (quartz tube ID 20 mm) to sublime 300 mg of starting material. The decrease of $T_{sblm}$ also reduces the concentration of heavier chemical impurities in the growth zone. Another important parameter is the purity of the starting material: the higher the initial purity, the less re-growth cycles are required. Using "sublimed grade" rubrene (Sigma-Aldrich), we were able to fabricate high-mobility devices after only 1 - 2 growth cycles.

The source, drain, and gate contacts were prepared by thermal evaporation of silver in high vacuum ($5\times10^{-7}$ Torr) on the crystal surface through a shadow mask. Typical in-plane channel dimensions were $L\times W$ = 1mm x1mm, where $L$ and $W$ are the channel length and width correspondingly. Earlier, we emphasized the importance of reducing the thermal load on a crystal during the deposition.[8] In order to achieve high mobilities, it is also crucial to avoid contamination of the channel surface by silver atoms deposited at oblique angles under the mask. Such contamination, which significantly deteriorates device performance, takes place due to the scattering of silver atoms from residual gas molecules. In order to prevent oblique angle deposition in the shadowed regions, we deposited silver through a "collimator", a narrow (4 mm ID) and long (30 mm) tube, positioned close to the crystal surface.

A thin film of parylene has been used as a gate insulator.[8] Prior to the deposition, the reactor was evacuated to a pressure of ~ 1 mTorr. The parylene deposition rate was ~ 300 A/min for the samples positioned ~ 35 cm away from the pyrolysis zone of the parylene reactor. The capacitance of the gate electrode per unit area, $C_i$, was measured for test capacitors fabricated simultaneously with the OFETs: $C_i$ = 2 ± 0.2 nF/cm$^2$ for a ~ 1-$\mu$m-thick parylene film. All the measurements reported in this paper have been done at room temperature in darkness using Keithley Source-Meters 2400 and Electrometers 617 and 6512. These measurements were taken at a slow gate voltage sweep with an average rate 2 V/min. No significant hysteresis or the time relaxation has been observed in these measurements.

Prior to the OFET fabrication, the bulk trap density ($N_t$) of the studied crystals has been estimated from charge transport measurements along the *c*-axis, in the direction perpendicular to the surface of thin crystals.[11] The crossover from the Ohmic regime to the space-charge-limited-current (SCLC) regime, and, for high purity crystals, to the trap-free (TF) regime can be observed with an increase of the bias voltage applied between two contacts deposited on the opposite facets of the crystal (Fig. 1). The presence of a linear Ohmic regime indicates that the non-linear contribution of Schottky barriers formed at the metal/rubrene interfaces is negligible in this case: the voltage

---
a) Electronic mail: podzorov@physics.rutgers.edu
b) Also at P. N. Lebedev Physics Institute, 119991 Moscow, Russia

drop across the Schottky barriers is small compared to the voltage drop across the highly resistive bulk crystal. Observation of the crossover to the SCLC regime at a low bias voltage $V_\Omega \sim 2.5$ V indicates that the charge carrier injection from the contacts is very efficient. From the threshold voltage of the TF regime, $V_{TF}$, a charge trap density $N_t$ as low as $10^{15}$ cm$^{-3}$ is estimated.[11]

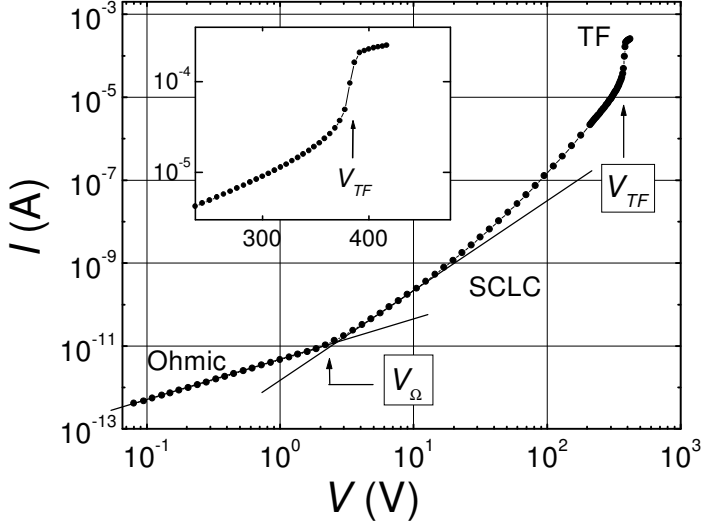

**FIG. 1.** *I-V* characteristic of a ~ 10 µm-thick rubrene crystal, measured along the c-axis. The inset is a blow-up showing the crossover to the trap free regime (also in a double-log scale).

We have studied two types of single-crystal OFETs: (a) conventional three-electrode devices and (b) devices with an additional pair of voltage probes located between the source and drain electrodes (the 2-probe and 4-probe configurations, respectively).[8] The study of 4-probe devices allowed us to measure separately the resistances of the conducting channel and the source and drain contacts, which is crucial for the extraction of the intrinsic charge carrier mobility, not limited by the contact resistance.

The trans-conductance characteristics of a rubrene *p*-type OFET in the 2-probe configuration are shown in Fig. 2. In these measurements, a positive voltage $V_{SD}$ is applied to the source electrode with respect to the grounded drain electrode.[12] The gate voltage, $V_g$, is applied to the gate electrode with respect to the drain electrode. The sharp increase of the source-drain current, $I_{SD}$, manifests a well-defined onset of the field effect, shown in Fig. 2 by the arrow for one of the source-drain voltages.

The field-effect onset has always occurred at a positive gate voltage, $V_g^{onset}$, similar to the data for well-ordered pentacene thin-film FETs.[7] This behavior resembles the operation of a 'normally-ON' FET with a built-in channel. The resemblance, however, is superficial: our observation that the sharp onset always occurs at $V_g^{onset} = V_{SD}$ indicates that the channel is induced electrostatically. Indeed, an application of a positive voltage $V_{SD}$ to the source electrode in the presence of the gate electrode only ~ 1 µm away from the interface creates a strong electric field normal to the crystal surface. This results in a propagation of the conducting channel from the source electrode to the drain at any $V_g < V_{SD}$. Such mode of operation indicates that the studied OFETs are essentially *zero threshold* devices.[13] The zero threshold operation suggests that the density of the charge traps is very low not only in the bulk, but also at the rubrene/parylene interface ($< 10^9$ cm$^{-2}$).[14]

The sharpness of the field-effect onset is usually characterized by the subthreshold slope, $S \equiv dV_g/d(\log I_{SD})$, which reflects the quality of insulator/semiconductor interfaces.[7] Since this quantity depends on the capacitance of the insulating layer $C_i$, it is convenient to introduce the normalized slope, $S_i \equiv SC_i$, in order to characterize the intrinsic properties of the semiconductor/insulator interface in various devices.

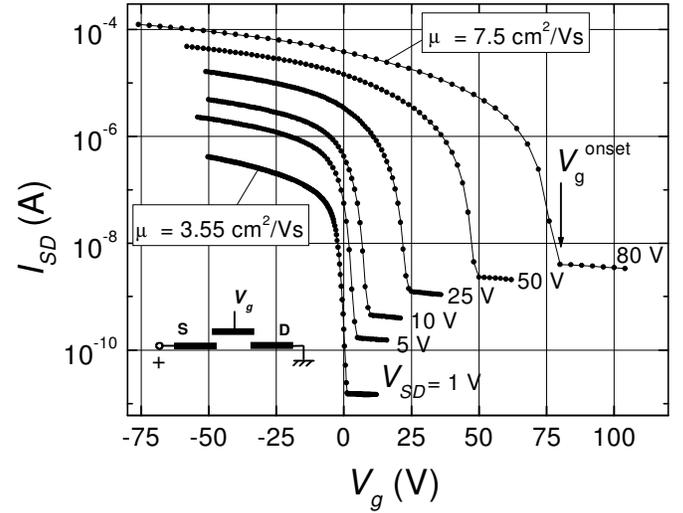

**FIG. 2.** The trans-conductance characteristics of an OFET fabricated on the rubrene single crystal, measured at different values of the source-drain voltage $V_{SD}$. The in-plane dimensions of the conducting channel are $L \times W = 1 \times 1$ mm$^2$.

Single-crystal OFETs exhibit a subthreshold slope as small as $S = 0.85$ V/decade, which corresponds to $S_i = 1.7$ V·nF/decade·cm$^2$. For comparison, the on/off transition in the best pentacene thin-film OFETs is much more extended: values of $S_i = 15 - 80$ V·nF/decade·cm$^2$ can be estimated from the data in Refs. [5, 7, 15]. We believe that the very small intrinsic subthreshold slope $S_i$ in the single-crystal devices reflects a high degree of ordering at the interface between the rubrene single crystal and parylene insulator. Single-crystal OFETs also compare favorably with α-Si:H FETs, for which $S_i \sim 10$ V·nF/decade·cm$^2$ has been reported.[3]

In the 2-probe measurements, the charge carrier mobility has been obtained from the trans-conductance characteristics using the expression $\mu = (L/WC_iV_{SD})(dI_{SD}/dV_g)$.[16] The typical dependence of the "2-probe" $\mu$ on the gate voltage is shown in Fig. 3. The maximum of $\mu$, observed at a small negative $V_g$, corresponds to "opening" of the Schottky barriers intrinsic to the source and drain contacts. The maxima of $\mu(V_g)$, as well as the dependence of $\mu$ on $V_{SD}$ (see Figs. 2, 3), are artifacts of the 2-probe measurements (see below). At sufficiently large negative gate voltage ($V_g \leq -20$ V), $\mu$ becomes almost $V_g$–independent (the highest variation observed is $\Delta\mu/\mu \sim 15$ %). This regime corresponds to the linear portion of the trans-conductance characteristics, where the conductivity of the channel, $\sigma = en_{2D}\mu$, varies almost linearly with the 2D carriers density $n_{2D}$. The maximum value of $\mu$ in the linear regime is ~ 8 cm$^2$/Vs.

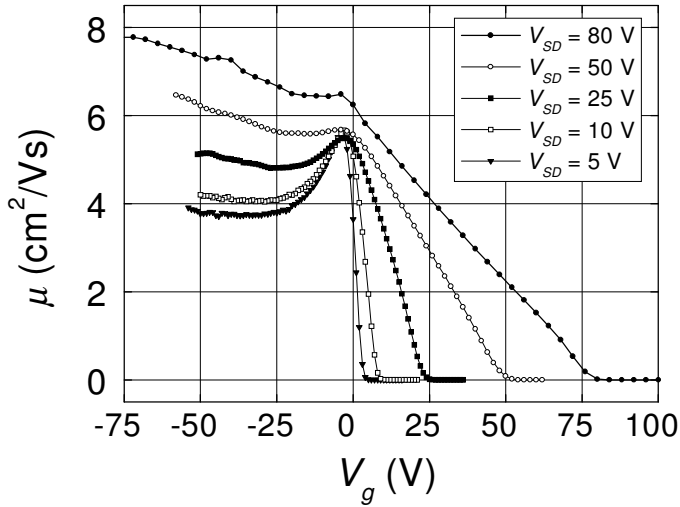

**FIG. 3.** The mobility $\mu = (L/WC_iV_{SD})(dI_{SD}/dV_g)$ versus the gate voltage, calculated from the data in Fig. 2.

With respect to the nearly $V_g$-independent mobility, the single-crystal OFETs compare favorably with the thin-film transistors, where a pronounced increase of $\mu$ with $V_g$ is observed due to the presence of structural defects.[5] Because of the strong $\mu(V_g)$ dependence in the thin-film OFETs, a large $|V_g| \geq 100$ V is often required to realize higher mobilities, comparable to that of $\alpha$-Si:H FETs (~ 0.5 cm$^2$/Vs).[3]

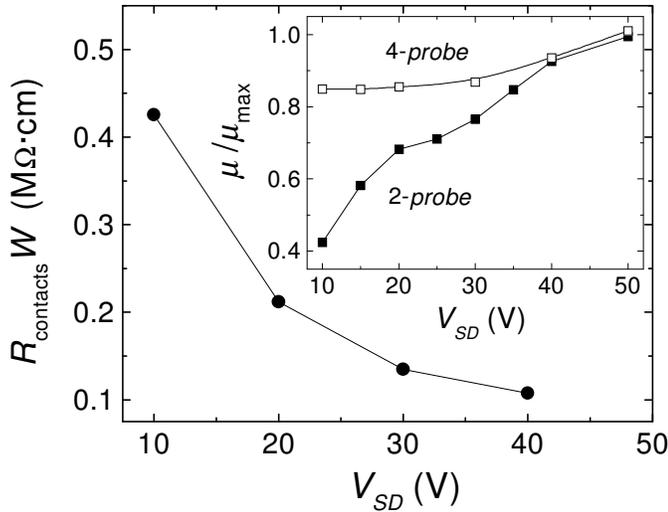

**FIG. 4.** The contact resistance normalized by the channel width $W$ measured for the 4-probe rubrene OFET as a function of the source-drain voltage ($V_g$ = -40 V). The inset: the mobility of the device measured in the 4-probe and 2-probe configurations, normalized by $\mu$ at $V_{SD} = 50$ V.

An apparent increase of the "2-probe" mobility with $V_{SD}$ (Figs. 2, 3) reflects a non-linear contribution of the Schottky barriers formed at the source and drain contacts.[17] In order to exclude the contact effects, we have measured the mobility and the contact resistance in the 4-probe configuration.[18] Figure 4 shows that the contact resistance decreases with $V_{SD}$. The dependences $\mu(V_{SD})$ measured in the 2- and 4-probe configurations are compared in the inset to Fig. 4. The "4-probe" data reflect the "intrinsic" charge carrier mobility, which is only weakly dependent on $V_{SD}$. The "2-probe" mobility converges to the "4-probe" mobility at high $V_{SD}$ due to decreasing contact resistance. This indicates that the "2-probe" mobility measured at sufficiently high source-drain bias is no longer limited by the contact resistance and, therefore, the values of $\mu$ obtained at high $V_{SD}$ are intrinsic (see Fig. 2).

In conclusion, by optimization of the fabrication process, we have significantly increased the field-effect mobility in rubrene-based single-crystal OFETs. In contrast to thin-film OFETs, single-crystal devices demonstrate zero threshold operation, gate-voltage-independent mobility and very small subthreshold slope. The single-crystal OFETs compare favorably not only with thin-film OFETs, but also with amorphous silicon transistors.

We thank Ch. Kloc for many useful discussions and Arnold Benenson for technical assistance. This work was supported in part by the NSF grant DMR-0077825, the DOD MURI grant DAAD19-99-1-0215, NATO grant PST.CLG.979275, INTAS (grant #01-2212), and RFBR (grant #03-02-16069).